\documentclass[aps,twocolumn,pre]{revtex4-1}

\usepackage{graphicx}
\usepackage{amsmath,mathtools}
\usepackage{color}

\begin{document}

\title{Nonequilibrium fluctuations in boson transport through squeezed reservoirs}

\author{Manash Jyoti Sarmah}
\author{Akanksha Bansal}
\author{Himangshu Prabal Goswami}
\email{hpg@gauhati.ac.in}
\affiliation{Department of Chemistry, Gauhati University, Jalukbari, Guwahati-781014, Assam, India}
\date{\today}

\begin{abstract} 
We investigate the effect of quantum mechanical squeezing on the nonequilibrium fluctuations of bosonic transport between two squeezed harmonic reservoirs and a two level system. A standard full counting statistics technique based on a quantum master equation is employed. We derive a nonzero thermodynamic affinity under equal temperature setting of two squeezed reservoirs.  The odd cumulants are shown to be independent of squeezing under symmetric conditions, whereas the even cumulants depend nonlinearly on the squeezing. The odd and even cumulants saturate at two different but unique values which are identified analytically. Further, squeezing always increases the magnitude of even cumulants in comparison to the unsqueezed case.  We also recover a standard steady state fluctuation theorem with a squeezing dependent thermodynamic affinity and  demonstrate the robustness of a steadystate thermodynamic uncertainty relationship.
\end{abstract}

\maketitle

\section{Introduction}
Squeezed states are specially designed non-canonical initial states which have a smaller variance in quadrature than that of a coherent state \cite{walls1983squeezed,schnabel2017squeezed,dodonov1991physical, puri1997coherent}. These quantum states form an exotic choice of initial states when studying atom-photon interactions in open quantum systems \cite{slusher1985observation,carmichael2011atom}. Squeezed states of light have been shown to exhibit non-trivial effects on system observables like quadrature autocorrelations in atom-cavity systems and higher order correlated photon pairs from  MgO:LiNbO$_3$ crystals. \cite{ourjoumtsev2011observation,mehmet2010observation}.
In studying open quantum systems from a quantum optics point of view, squeezed light is usually modeled as a harmonic reservoir of photonic modes by introducing an additional  parameter to allow quantum control \cite{dodonov2002nonclassical,lutkenhaus1998mimicking,banerjee2008geometric}. Quantum control of matter properties through squeezed harmonic reservoirs  is not limited to optics and has also been studied with respect to quantum thermodynamics and information theories \cite{manzano2016entropy,li2017production,you2018entropy,jahromi2020quantum,macchiavello2020quantum,zou2022geometrical}. 
In open quantum systems that are out of equilibrium, it has been shown that,
using  squeezed or correlated thermal reservoirs, the power and efficiency of heat engines can be improved, even surpassing the Carnot bound \cite{manzano2016entropy,niedenzu2016operation,agarwalla2017quantum,klaers2017squeezed,newman2017performance}. In quantum information theories, it has been reported that squeezed reservoirs lack a proper description of temperature due to mutual information exchanged between system and reservoirs via manifestation of excess heat or work which is addressable through mutual von Newmann entropy production rates \cite{manzano2016entropy,li2017production,you2018entropy}.

With respect to nonequilibrium quantum thermodynamics, the understanding of quantum signatures is murky due to the rich dynamics present in the form of higher order fluctuations \cite{wang2021nonequilibrium,hsiang2021nonequilibrium}. Such  statistical features manifest into a general set of mathematical relationships between the forward and backward probabilities of exchanged quantities (particle or energy) called fluctuation theorems (FT) and inequalities between output power and noise called thermodynamic uncertainty relationships (TUR) \cite{NatPhys.16.15,PhysRevB.98.155438,pietzonka2017finite}. FT ensures validity of detailed balance during particle or energy exchange between system and reservoirs while the TUR represents a trade-off between power and noise.  Quantum signatures in fluctuation theorems have been theoretically studied to a great extent and even experimentally realized \cite{perarnau2019quantum,hernandez2021experimental}. Squeezed states along with coherent and cat states are shown to generate additional corrective parameters on the classical fluctuation theorem of the Crooks type \cite{holmes2019coherent}. Very recently, it has been shown that noncommutive initial energy measurements with the squeezed reservoirs may lead to negative probabilities and short time measurements and also may not lead to a Jarzynski-Wojcik type of fluctuation theorem \cite{yadalam2022counting}. Possibilities are still being explored to  garner a general understanding of the role of squeezed initial states in FTs and TURs \cite{huang2012effects,hsiang2021fluctuation,talkner2013statistics}. It has been recently reported that the higher order fluctuations during photon transport can be maximized due to mixing between a qubit and squeezed resonators \cite{wang2021nonequilibrium}. In general however, the fluctuations of particle (boson or fermion) transfer in squeezed nonequilibrium systems are not yet fully understood  demanding further studies \cite{chen2022nonclassical}. 

In this work we study the effect of squeezing on the statistical fluctuations in bosonic transport through two squeezed thermal reservoirs and assess the validity of established FT and TUR. Such models are pretty standard and well studied in quantum transport \cite{giraldi2014coherence,PhysRevLett.104.170601,pekola2021colloquium}. In this model, in absence of squeezing, a steadystate FT of the type, $\ln[(P(q)/P(-q)] = q\Delta\beta$ \cite{saito2007fluctuation,PhysRevLett.104.170601}, exists  where $q$ is the net number of particles exchanged between system and reservoir while $\Delta\beta$ represents the temperature gradient between the two reservoirs and is directly proportional to the thermodynamic affinity ($A_o$) that drives the system out of equilibrium. This affinity is related to the noise (second order fluctuation) via a proposed steadystate TUR, $FA_o\ge 2k_B$, where $F$ is the Fano factor. 
FTs are direct consequence of Tasaki-Crooks or Gallavoti-Cohen type of symmetry in the characteristic functions of boson exchange statistics \cite{hertz1910uber,esposito2009nonequilibrium}. It has been argued that, when the system has microcanonical initial states,   the
characteristic functions do not exhibit such a symmetry \cite{talkner2013statistics} due to the fact that the initial conditions for
the forward and the backward processes are different. 
 Our work focuses on assessing the validity of such relationships via a quantum statistical approach by employing the well established methodology of full counting statistics (FCS) \cite{esposito2009nonequilibrium} within a quantum master equation framework. This paper is organized as follows. Firstly, in Sec.(\ref{model}) we present our model and the general formalism used. In Sec.(\ref{results}), we show our results and analysis after which we conclude in Sec. (\ref{Conclusion}).

\section{Model and Formalism}
\label{model}
\begin{figure}
\centering
\includegraphics[width = 0.48 \textwidth]{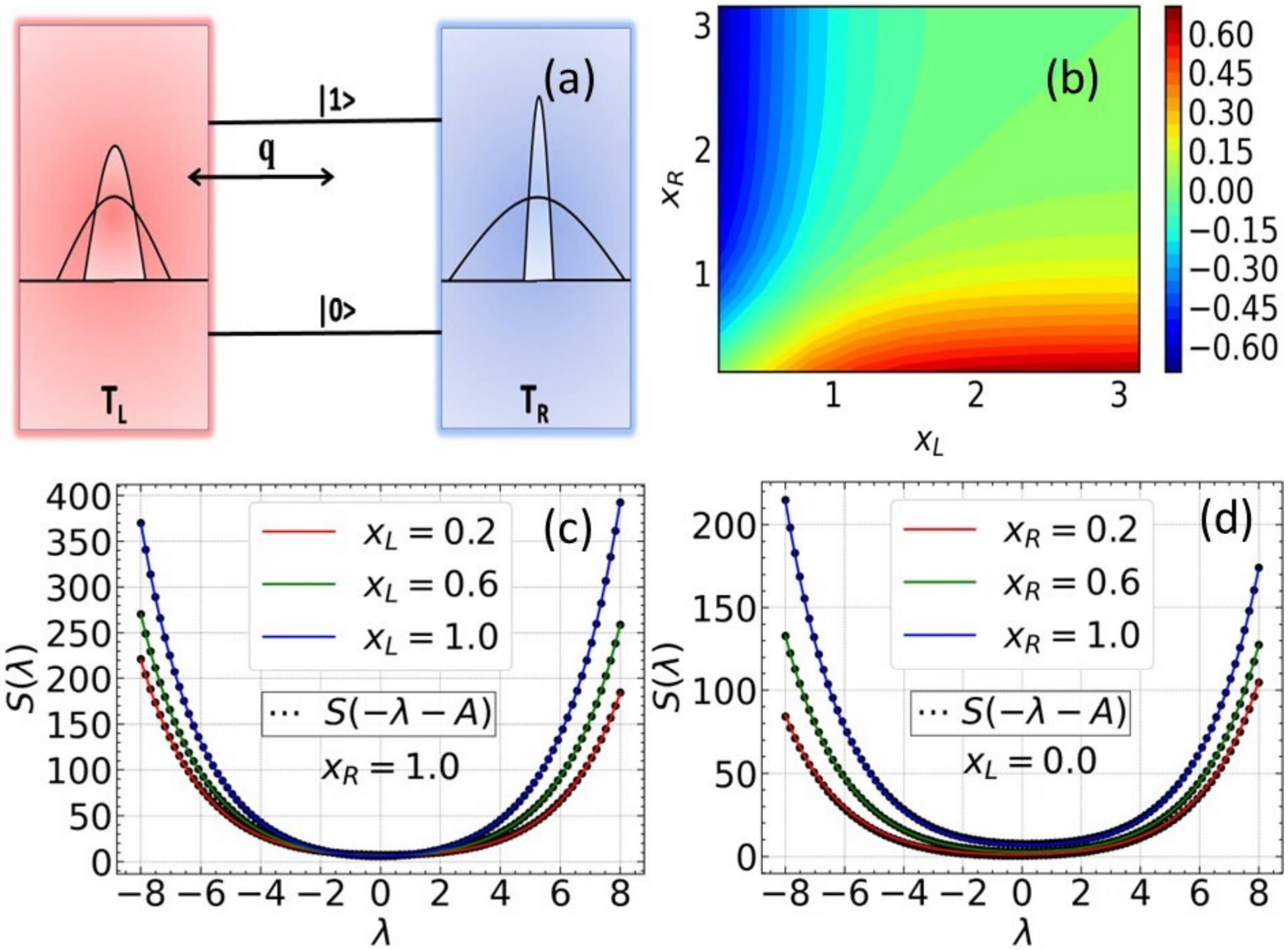}
\caption{a) Schematic diagram of two squeezed harmonic baths interacting with  two energy levels. b) Density plot of the thermodynamic affinity under equal temperature setting as per Eq.(\ref{aff-eq}) with $\omega_o=0.5 (\hbar=1,k_B=1), T_L=T_R=0.7$. In (c) (d) the  GC symmetry is shown as per Eq.(\ref{eq-GC}) with $T_L=0.7,T_R=0.4$. The solid (dotted) line represent $S(\lambda)(S(-\lambda-A))$. Throughout the numerics, we have used $\Gamma_L = \Gamma_R=1s^{-1}$.
}
\label{schematic}
\end{figure}
The Hamiltonian of the two level system interacting with two bosonic reservoirs (Fig.(\ref{schematic})a) can be written as,
\begin{align}
\label{ham-eq}
    \hat{H}&=\displaystyle\hbar\omega_o\hat{b}^\dag\hat{b}+\sum_{i=\nu,\nu\in L,R}^{}\hbar\omega_i^{}\hat{a}_i^\dag\hat{a}_i^{}+\hat{V},
\end{align}
with
\begin{align}
       \label{V-def}
    \hat{V}&=\sum_{i,\nu\in L,R}k_{i}^\nu(\hat{a}_{i\nu}^\dag\hat{b}+\hat{a}_{i\nu}\hat{b}^\dag)
\end{align}
Here, $\hbar\omega_o\hat{b}^\dag\hat{b}$ is the on site Hamiltonian with bare frequency $\omega_o$, while $\hat{b}^\dag(\hat{b})$ is the bosonic creation (annihilation operator) on the site. The second term is the reservoir Hamiltonian with squeezed harmonic states and is a sum of two terms that represent the left (L) and right (R) squeezed reservoirs.  The single particle operators $\hat{a}_{i\nu}^\dag(\hat{a}_{i\nu})$ represent the creation (annihilation) of a boson in  the i-th mode from (of) the $\nu$-th bath. $\hat{V}$ is the system bath coupling Hamiltonian with  $k_i ^ \nu$ being the coupling constant for the i-th squeezed mode of the $\nu$th bath to the bare site mode. The schematic representation of the model is shown in Fig.(\ref{schematic} a). The squeezed density matrix for the $\nu$-th reservoir ($\hat H_\nu$ being the $\nu$th reservoir Hamiltonian) is given by
\begin{align}
    \hat\rho_\nu&=\frac{1}{Z} \exp\{-\beta_\nu^{} \hat{S}\hat H_\nu\hat S^\dag\},
\end{align}
with $\beta_\nu = (k_B T_\nu)^{-1}$ being the inverse temperature and $\hat S$ is the squeezing operator on the $k-th$ bath mode given by :
\begin{align}
    \hat S &=\displaystyle \prod_{k} e^{\frac{1}{2}(\lambda_{k\nu}^* \hat a_{k\nu}^{\dag2}-h.c)},\\
    \lambda_{k\nu} &= x_{k\nu} e^{i\theta_{k\nu}} , x_{k\nu}>0 
\end{align}
with $x_{k\nu},\theta_{k\nu}$ being the squeezing parameters \cite{dodonov2002nonclassical,li2017production,yadalam2022counting}.

We are interested in the statistics of boson transport between system, represented by the reduced density matrix $\hat\rho$ and the squeezed baths. We choose to count the net number of bosons exchanged with the left reservoir and denote it as $q$. To keep track of $q$, we can write down a moment generating vector for the reduced system  within the FCS formalism \cite{esposito2009nonequilibrium,harbola2007statistics,goswami2015electron}. We formulate this in the Liouville space in terms of the auxiliary counting field, $\lambda$. The reduced moment generating density vector, $|\breve\rho(\lambda,t)\rangle\rangle$, can be written as,
\begin{equation}
\label{lvl-eqn}
    |\dot{\breve\rho}(\lambda,t)\rangle\rangle  =   {\breve{\cal L}} (\lambda)|\breve \rho(0,0)\rangle\rangle,
\end{equation}
where the elements of the density vector contains the populations of the occupied and unoccupied states, $\{\rho_{11}$, $\rho_{00}\}$ (appendix) with the evolution superoperator, ${\breve{\cal L}} (\lambda)$, given by
\begin{align}
\label{Louv-eq}
{\breve{\cal L}} (\lambda)
&=
\left[\begin{array}{cc}
    -\alpha_L-\alpha_R & \beta_Le^{\lambda}+\beta_R \\
    \alpha_Le^{-\lambda}+\alpha_R & -\beta_L-\beta_R
\end{array}\right].
\end{align}
The rates of boson exchange between system and reservoirs are given by (we ignore the Lamb shifts terms):
\begin{align}
    \alpha_\nu&=\Gamma_\nu(1+N_\nu), \nu=L,R
\\
\beta_\nu&=\Gamma_\nu N_\nu
\end{align}
with the occupation factor being
\begin{align}
    N_\nu&=\cosh{(2x_\nu)\bigg{(}n_\nu^{}+\frac{1}{2}\bigg{)}-\frac{1}{2}}
    \end{align}
    where  $n_\nu=(e^{\beta_\nu\hbar\omega_o}-1)^{-1}$
being the the Bose-Einstein distribution of the $\nu$-th bath. $x_\nu >0$ is the renormalized parameter responsible for squeezing the $\nu$-th harmonic bath within the Markovian regime \cite{li2017production} (see the appendix).  In the steadystate, one of smaller eigenvalues of the RHS of Eq.(\ref{Louv-eq}), $S(\lambda)$ corresponds to a cumulant generating function, from which the cumulants of boson exchange can be directly evaluated as,
\begin{align}
\label{cum-eq}
 j^{(n)}&=\partial_\lambda^n S(\lambda)|_{\lambda=0}.
\end{align}
Here, $n=1,2,3$ and $4$ corresponds to the flux, noise, skewness and kurtosis respectively, which we evaluate in the next section. 
\begin{figure}[bth!]
\centering
\includegraphics[width = 0.48 \textwidth]{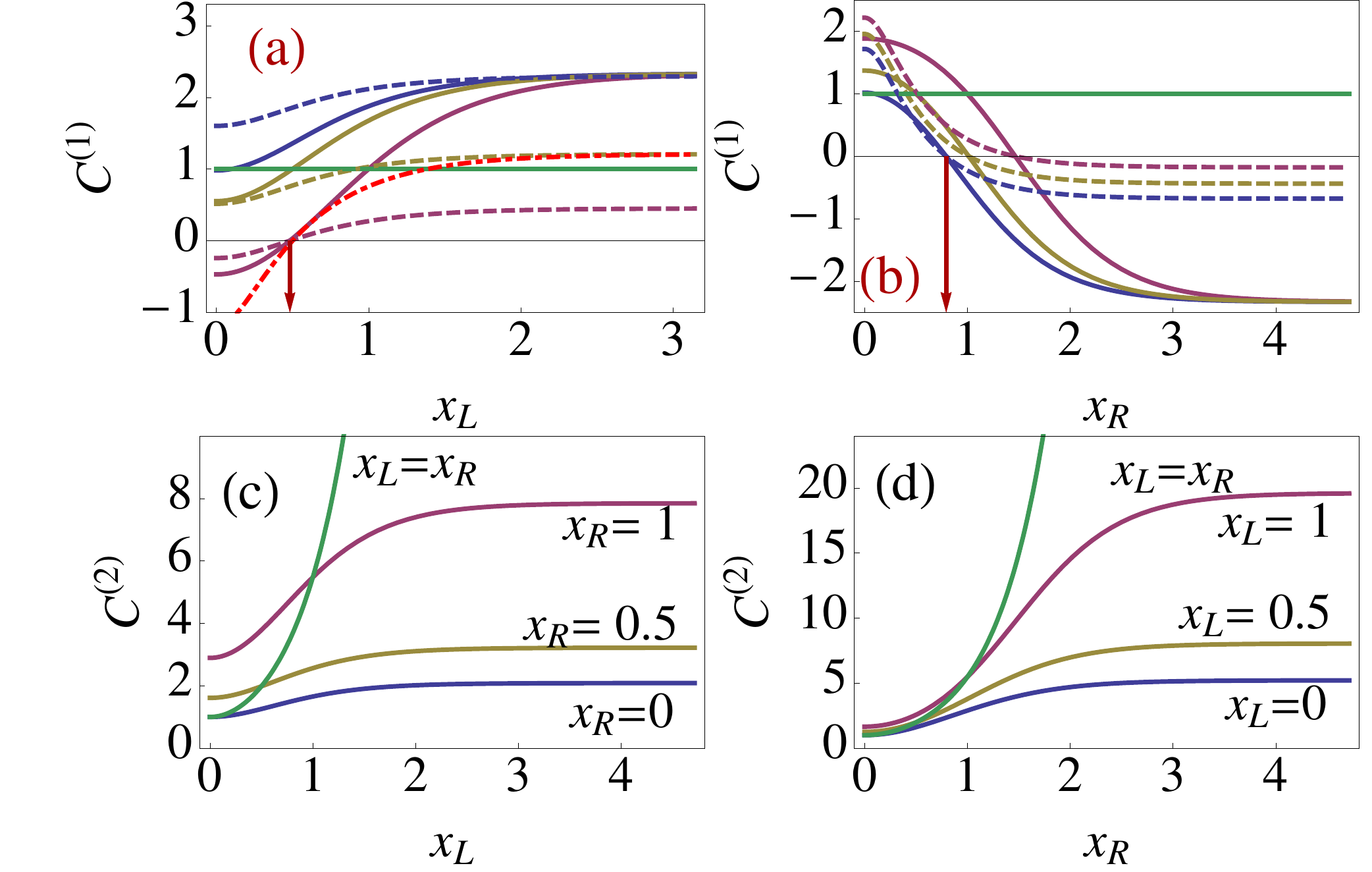}
\caption{(Color online) Plots showing the behavior of the flux ($C^{(1)})$ and noise, ($C^{(2)})$) as a function of the squeezing parameters $x_L$ and $x_R$ with $n_L=1$ and $n_R=0.1$.  In (a) and (b), the solid lines represent flux and the dotted lines represent $A$. The straight line at unity (green) is evaluated at $x_L=x_R$ and the other curves are evaluated at $x_\nu$ =1 (purple), 0.5(golden) and 0(blue). Both the quantities saturate at high squeezing values. The saturation value of A as a function of $x_L(x_R)$ is dictated by Eq. (\ref{AlimL-eq}(\ref{AlimR-eq})). The saturation value of $C^{(1)}$ as a function of $x_\nu$ is at $\Gamma_\nu\tilde n_\nu\cosh(2x_\nu)/j^{(1)}_o$. The change in the direction of flux occurs at $x_\nu^*,\nu=L,R$ (Eq.(\ref{switchL-eq})) and is indicated by the downward pointing arrow. In (a), the red dotted curve represents $A$ evaluated for $x_R=0.5 (n_L=n_R)$. In subfigures, (c) and (d) the noise is seen to be always greater than unity and saturates at $\Gamma_\nu\tilde n_\nu/j^{(2)}_o.$ 
}
\label{fig-flux-ratio}
\end{figure}


\section{Results and Discussion}
\label{results}
We start our discussion by stating that, the analytical form of the cumulant generating function $S(\lambda)$ is given by
\begin{align}
\label{S-lam-eq}
    S(\lambda)=&-(\alpha_L+\alpha_R+\beta_L+\beta_R)\\&+\sqrt{(\alpha_L+\alpha_R+\beta_L+\beta_R)^2+4 f(\lambda)}) \\
    f(\lambda)&=\alpha_L\beta_R(e^{-\lambda}-1)+\alpha_R\beta_L(e^\lambda-1)
\end{align}
 and has the same mathematical structure as the one obtained from harmonic baths \cite{PhysRevB.98.155438,PhysRevLett.104.170601}, albeit with squeezed rates. Hence it satisfies a Gallavoti-Cohen (GC) symmetry 
 \begin{align}
  \label{eq-GC}
 S(\lambda)=S(-\lambda-A) 
 \end{align}
  with a squeezing dependent thermodynamic affinity, A, given by,
 \begin{equation}
 \label{aff-eq}
 A=\log \left(\frac{(\tilde n_L^{} \cosh (2 x_L^{})-1) (\tilde n_R^{} \cosh (2 x_R^{})+1)}{(\tilde n_L^{} \cosh (2
   x_L^{})+1) (\tilde n_R^{} \cosh (2 x_R^{})-1)}\right)
 \end{equation}
where we have defined $\tilde n_\nu=1+2n_\nu$. The GC symmetry ensures the validity of a steadystate FT of the type, $P(q)/P(-q)=\exp(qA)$. The validity of GC symmetry is graphically shown in Fig.(\ref{model}b,c) for various squeezing values. Note that, one recovers the standard expression of the affinity (we denote it as $A_o$), $A_o=\Delta\beta=(\hbar\omega_o/k_B)(T_L^{-1}-T_R^{-1})/k_B$ when $x_\nu=0$. Note that, when $T_L=T_R$, the affinity goes to zero for the unsqueezed case (harmonic case) and indicates thermal equilibrium ($A_o=0$). However, for the squeezed case, maintaining $T_L=T_R$ in Eq.(\ref{aff-eq}),  we find that $A\ne 0$ as long as $x_L\ne x_R$. Thus by unsymmetrically squeezing the two reservoirs, one can kick the system out of equilibrium even when the temperatures are  equal. We graphically show the nonzero value of A under the equal temperature setting in Fig.(\ref{model}d) by varying both $x_L$ and $x_R$.  This is graphically also shown in the dashed and dotted curve in Fig.(\ref{fig-flux-ratio}a). In Fig.(\ref{fig-flux-ratio} a (b)) each of the dashed curves represent the thermodynamic affinity evaluated at different $x_R (x_L)$ as a function of $x_L (x_R)$. These curves saturate for high squeezing values. These saturation values are given by,
\begin{align}
\label{AlimL-eq}
 \lim_{x_L\to\infty}A&=\log\frac{\tilde n_R\cosh(2x_R^{})+1}{\tilde n_R\cosh(2x_R^{})-1}\\
 \label{AlimR-eq}
 \lim_{x_R\to\infty}A&=\log\frac{\tilde n_L\cosh(2x_L^{})-1}{\tilde n_L\cosh(2x_L^{})+1}.
 \end{align}
The above two equations indicate that, only one value of the Bose-Einstein factor dominates in the high squeezing limit. Both the saturated values are finite and greater than zero even when $T_L=T_R (n_L=n_R)$. The saturated values become zero if and only if $x_L=x_R$ as well as $n_L=n_R$.  


\begin{figure}[bht!]
\centering
\includegraphics[width = 0.48 \textwidth]{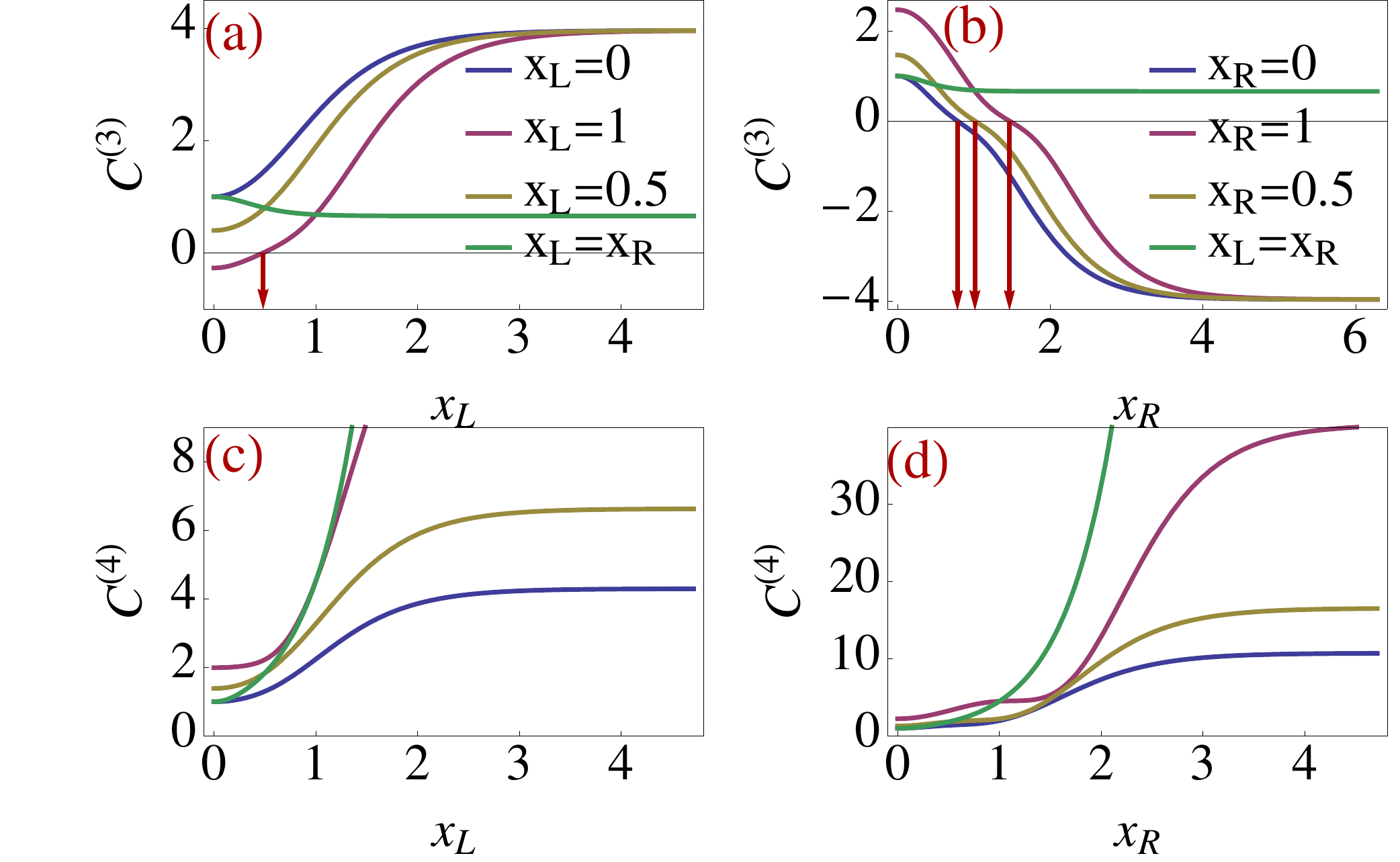}
\caption{(Color online) Plots showing the behavior of the skewness ($C^{(3)})$ and kurtosis, ($C^{(4)})$) as a function of the squeezing parameters.  In (a) and (b), the solid lines represent flux and the dotted lines represent $A$. The straight line at unity (green) is evaluated at $x_L=x_R$ and the other curves are evaluated at $x_\nu$ =1 (purple), 0.5(golden) and 0(blue).  In subfigures, (c) and (d) the kurtosis  is seen to be always greater than unity. Both cumulants saturate at the values predicted by Eqs.(\ref{sat-even-eq},\ref{sat-odd-eq}). Numerical parameters used are given in Fig.(\ref{fig-flux-ratio}) and color-matched.}

\label{fig-fluc-ratio}
\end{figure}

We can now proceed to evaluate the cumulants using Eqs.(\ref{Louv-eq}) and (\ref{cum-eq}). Note that, one recovers the standard generating function for bosonic transport in absence of squeezing,  $x_\nu=0$ in Eq.(\ref{Louv-eq})\cite{PhysRevB.98.155438}. 
We denote the cumulants in absence of squeezing as $ j^{(n)}_o$ and define a dimensionless ratio representing the scaled cumulants,
\begin{align}
 \label{dim-eq}
 C^{(n)}:=\displaystyle\frac{ j^{(n)}}{ j^{(n)}_o}
\end{align}
which signifies the extent of change in the values  of the squeezed cumulants in comparison to unsqueezed case. When$|C^{(n)}|>(<)1$, the squeezing increases (decreases) the value of the cumulant in comparison to the unsqueezed case. We plot the dimensionless cumulant ratio $C^{(n)}$ as a function of squeezing parameters which are shown graphically in figures (\ref{fig-flux-ratio}) and (\ref{fig-fluc-ratio}) for $n=1,2,3$ and $4$.

The first cumulant of flux can be be written as
\begin{align}
\label{flux-eq}
 j^{(1)}&=\frac{\Gamma (\tilde{n}_L \cosh (2 x_L^{})-\tilde{n}_R \cosh
   (2 x_R^{}))}{\Gamma_L^{} \tilde{n}_L \cosh (2 x_L^{})+\Gamma_R^{}\tilde{n}_R^{} \cosh (2 x_R^{})}
\end{align}
with $\Gamma = \Gamma_L\Gamma_R$. Under the symmetric squeezing case, $x_L=x_R$, the above equation reduces to the standard expression of flux in absence of squeezing,
\begin{align}
\label{flux-0-eq}
  j^{(1)}\big{|}_{x_L=x_R}= j^{(1)}_o=\frac{2\Gamma(n_L^{}-n_{R})}{\Gamma_L\tilde n_{L}+\Gamma_R\tilde n_R},
\end{align}
From the above equation, it is evident  that the flux is independent of squeezing when both the bosonic reservoirs are squeezed symmetrically. The quantity $C^{(1)}$ is graphically shown in Fig.(\ref{fig-flux-ratio}a,b). The crossover from negative to positive flux happens since the thermodynamic affinity becomes negative (shown as dotted lines in Fig.(\ref{fig-flux-ratio}a,b)). This switching of the direction of flux occurs at the point $x_{\nu}^*$
\begin{align}
 \label{switchL-eq}
 x_{\nu}^*=\frac{1}{2}&\cosh ^{-1}\left(\frac{\tilde n_\nu^{} \cosh (2x_\nu^{})}{\tilde n_\nu^{'}}\right),
 \nu\ne\nu'\in L,R
\end{align}
and shown graphically via  downward arrow in Fig.(\ref{fig-flux-ratio}a,b). Further, whenever  $\cosh(2x_L)>(<)\cosh(2x_R)$, $C^{(1)}>(<)1,$ if $T_L\ge T_R$ is maintained. 

\begin{figure}
\centering
\includegraphics[width= 0.47 \textwidth]{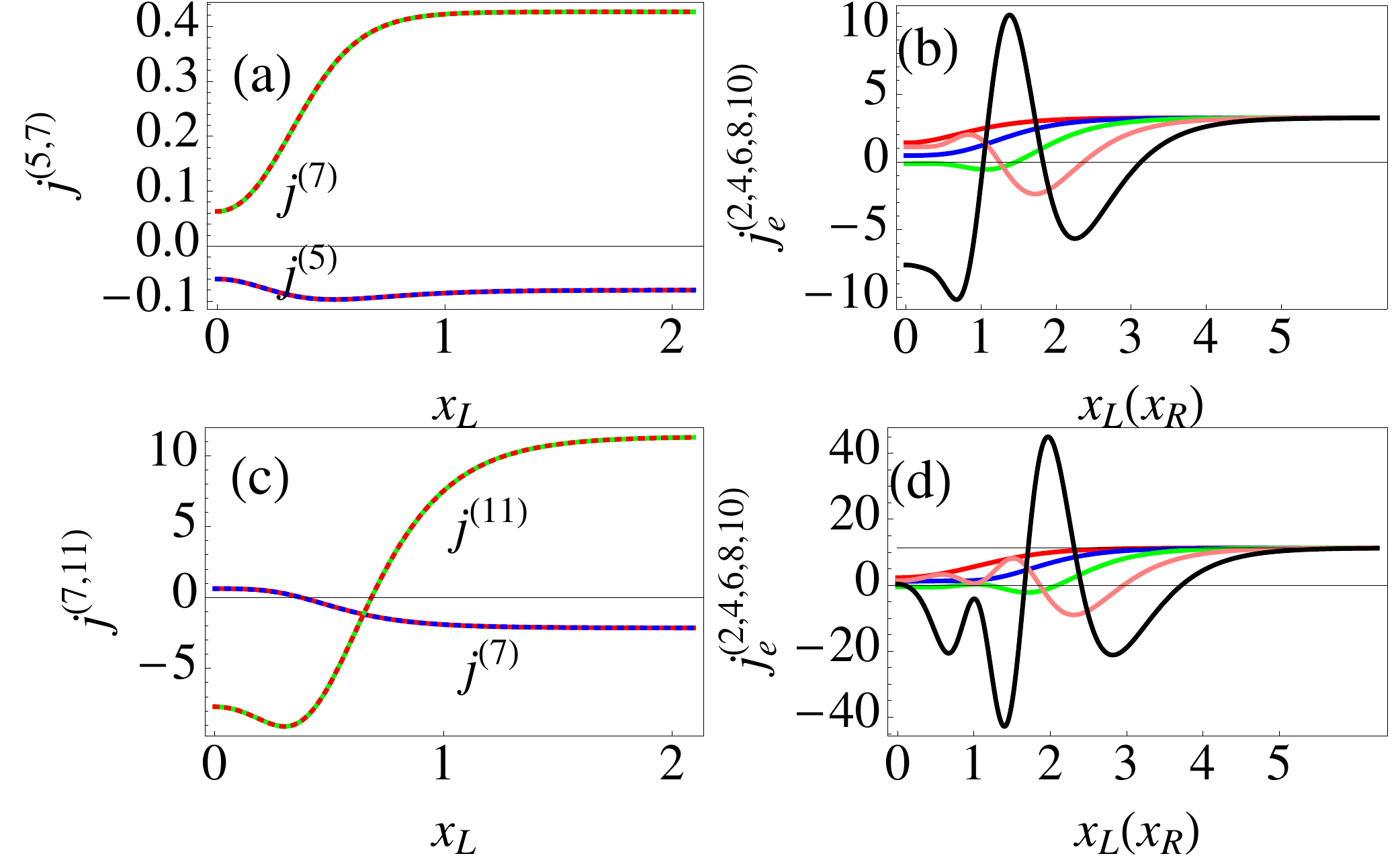}
\caption{(Color Online) Behavior of higher odd and even cumulants. In (a) and (c), the values of odd cumulants are same for symmetric squeezing and unsqueezed case. These saturate to the scaled values given by Eq.\ref{sat-odd-eq}) In (b), $x_\nu=0.2$ and (d)$x_\nu=1$,  the even cumulants are evaluated at $T_L=T_R$ which saturate at the values dictated by Eq.(\ref{sat-even-eq}), indicated by straight line in (d). The red curves represent the analytical result from Eq.(\ref{eq-j2-sym}). The other curves are numerically evaluated for $n =4, 6, 8 $ and 10 shown by the  blue, green, pink and  black curves, respectively. }
\label{fig-high-cum}
\end{figure}

The noise or second order cumulant can be expressed as,
\begin{widetext}
\begin{align}
\label{fluc-eq}
j^{(2)}&=\frac{\Gamma \left((\tilde{n}_L \tilde{n}_R \cosh (2 x_L^{})
   \cosh (2 x_R^{})-1) (\Gamma_L^{} \tilde{n}_L \cosh (2 x_L^{})+\Gamma_R^{}\tilde{n}_R \cosh (2 x_R^{}))^2-\Gamma (\tilde{n}_L \cosh
   (2 x_L^{})-\tilde{n}_R \cosh (2 x_R^{}))^2\right)}{(\Gamma_L
   \tilde{n}_L \cosh (2 x_L^{})+\Gamma_R\tilde{n}_R \cosh (2
   x_R^{}))^3}
   \end{align}
 which under equal squeezing ($x_L=x_R$) can be recast as: 
 \begin{align}
  \label{fluc-sym-eq}
  j^{(2)}|_{x_L\to x_R}^{}&=\frac{-\Gamma ((\Gamma_L^{} + \Gamma_R^{}) (\Gamma_L \tilde n_L^{2} + \Gamma_R\tilde n_R^{2} ) + \tilde n_L^{} \tilde n_R^{} (\Gamma_L\tilde n_{L}^{} + \Gamma_R\tilde n_{R})^2 \cosh^2(2x_L)) \text{sech}(
  2 x_R)}{(\Gamma_L\tilde n_L^{}  + \Gamma_R \tilde n_R^{})^3},
 \end{align}

and reduces to the standard known expression for the second cumulant \cite{saryal2019thermodynamic} only when, $x_R=x_L$=0, 
\begin{align}
\label{fluc-0-eq}
 j^{(2)}\big{|}_{x_L,x_R=0}&=j^{(2)}_o=\frac{2 \Gamma (2 n_L^{} n_R^{}+n_L^{}+n_R^{})}{\Gamma_L^{} \tilde n_L^{}+\Gamma_R^{}\tilde n_R^{}}-\frac{4\Gamma^2 (n_L^{} -  n_R^{})^2}{(\Gamma_L^{} \tilde n_L^{}+\Gamma_R\tilde n_R^{})^3}
\end{align}
\end{widetext}
Note that under symmetric condition $x_L=x_R$, the RHS of Eq.(\ref{fluc-sym-eq}) is not equal to the RHS of Eq.(\ref{fluc-0-eq}) indicating the dependence of noise under symmetric conditions unlike the flux (Eq.(\ref{flux-0-eq})). The quantity $C^{(2)}$ is shown graphically as a function of squeezing parameters in Fig.(\ref{fig-flux-ratio}c,d). Note that, $C^{(2)}>1$ is maintained for all values of $x_L,x_R>0$, indicating that squeezing always increases the magnitude of the noise. 
The skewness ($n=3$) can be expressed as (for compactness, we have used a shorthand notation, $\tilde n_\nu(x_\nu):=1+2n_\nu\cosh(2x_\nu)$):
\begin{widetext}
\begin{align}
    \label{skew}
    j^{(3)}&=\displaystyle\frac{{\Gamma (\tilde{n}_L(x_L^{})-\tilde{n}_R (x_R^{})) \left(\Gamma_L^4 \tilde{n}_L^4 (x_L)+\Gamma_L^3\Gamma_R \tilde{n}_L^3(x_L^{}) \tilde{n}_R^3 (x_L^{})+3 \Gamma_L^2\Gamma_R  (\Gamma_L^{}+\Gamma_R) \tilde{n}_L^2(x_L)\atop
   +\frac{ 1}{2} \Gamma_R^2\tilde{n}_R^2(x_R^{})  \left(6 \Gamma_L^2+6 \Gamma+\Gamma_R^{2}\tilde n_R^{2}(2x_R)+\Gamma_R^2\tilde{ n}_R^{2}\right)+\Gamma_L\Gamma_R^3 \tilde{n}_L(x_L^{}) \tilde n_R^3(
   x_R^{})\right)}}{(\Gamma_L \tilde{n}_L (x_L^{})+\Gamma_R^{}\tilde n_R^{}(x_R))^5}
   \end{align}
    which under the symmetric squeezing ($x_L=x_R$) is independent of squeezing and reduces to the standard result\cite{saryal2019thermodynamic},
\begin{align}
 j^{(3)}\big{|}_{x_L=x_R}&=j^{(3)}_o=\frac{2 \Gamma (n_L^{}-n_R^{}) \left((\Gamma_L^{} \tilde n_L^{}+\Gamma_R\tilde n_R^{})^4-6 \Gamma(2 n_L^{} n_R^{}+n_L^{}+n_R^{}) (\Gamma_L^{}
   \tilde n_L^{}+\Gamma_R^{} \tilde n_R^{})^2+12\Gamma (n_L^{}- n_R^{})^2\right)}{(\Gamma_L^{} \tilde n_L^{}+\Gamma_R\tilde n_R^{})^5}
\end{align}
   \end{widetext}
  The quantity $C^{(3)}$ is graphically shown as a function of squeezing parameters in Fig.(\ref{fig-fluc-ratio}a,b). Note that, similar to the flux, a switching in the sign of the scaled skewness is observed as one keeps increasing the squeezing. 
The behavior of switching of the skewness from negative to positive can be understood in a similar manner to that of the flux. The switching point is given by Eq.(\ref{switchL-eq}) and are indicated by the downward arrows in Fig.(\ref{fig-fluc-ratio}a,b). $C^{(3)}>(<)1$ as long as $\cosh(2x_L)>(<)\cosh(2x_R)$ if $n_L>(<)n_R$.  
We conclude that all odd cumulants are independent of squeezing when both the reservoirs are squeezed symmetrically. This is graphically shown in Fig.(\ref{fig-high-cum}a,c) by numerically evaluating the odd cumulants. As can be seen from the figure,  the dotted curves that represent $j^{(n)}_o$,   overlap with the solid curves, $j^{(n)}$ for the cumulants $n=5,7,9$ and $11$. The even cumulants however depend on the squeezing under symmetric conditions. 
We also find that all the odd and the even cumulants saturate to two different unique values, $j_{\infty}$ as a function of either of the squeezing parameters. The value of odd cumulants under large squeezing values is given by,
\begin{align}
 \label{sat-odd-eq}
 \lim_{x_{\nu'}\to\infty}j^{(n)}|_{n\in odd}=j_{\infty}^{odd}=\Gamma_\nu^{}\tilde{n}_\nu^{}\cosh(2x_\nu^{}), \nu\ne\nu' \in L,R,
\end{align}
while the even cumulants' saturation value is obtained at
\begin{align}
 \label{sat-even-eq}
 \lim_{x_{\nu'}\to\infty}j^{(n)}|_{n\in even}=j_{\infty}^{even}=\Gamma_\nu, \nu\ne \nu' \in L,R,
\end{align}
These saturation values are highlighted in Fig.(\ref{fig-high-cum}) for $n=2\ldots 12$. The saturation behavior are also visible in figures (\ref{fig-flux-ratio} and \ref{fig-fluc-ratio}), albeit scaled as $j^{(n)}_\infty/j^{(n)}_o$. An application of this result is that, Eq.(\ref{sat-even-eq}) together with Eq.(\ref{sat-odd-eq}) would allow experimental determination of the coupling values and the Bose- Einstein distribution by simply squeezing the system to a large extent.

In Fig.(\ref{fig-high-cum} b and d), we show the even cumulants ($n=2,
\ldots 10$) as a function of squeezing parameters under the condition $T_L=T_R$ and $\Gamma_L=\Gamma_R$. Under this case, we find that the cumulants behave identically when $x_\nu$ is varied for a fixed $x_{\nu'}$. We explain this by taking the analytical expression of the noise in the aforesaid limit which is given by,
\begin{align}
\label{eq-j2-sym}
j^{(2)}_e&=j^{(2)}|_{\Gamma_L\to\Gamma_R,n_L\to n_R}\\ \nonumber
&=\displaystyle\frac{\Gamma_L \left(\tilde n_R^{2} \cosh ^3(2 x_L^{}) \cosh (2 x_R^{})+\tilde n_R^{2} \cosh (2
   x_L^{}) \cosh ^3(2 x_R^{})\atop+2 \cosh ^2(2 x_L^{}) \left(\tilde n_R^{2} \cosh ^2(2
   x_R^{})-1\right)-2 \cosh ^2(2 x_R^{})\right)}{\tilde n_R^{} (\cosh (2 x_L^{})+\cosh
   (2 x_R^{}))^3}.
\end{align}
The above expression is symmetric with respect to the exchange of  $x_L$ with $x_R$ and vice-versa and hence results in an equivalent dependence of the even cumulants on the squeezing parameters. Note that, the thermodynamic affinity, $A$, is however antisymmetric with respect to exchange of $x_L$ and $x_R$ under the symmetric condition, i.e, $A(x_L;x_R)=-A(x_R;x_L)$, as evident from Eq.(\ref{aff-eq}).

\begin{figure}
\centering
\includegraphics[width = 0.48 \textwidth]{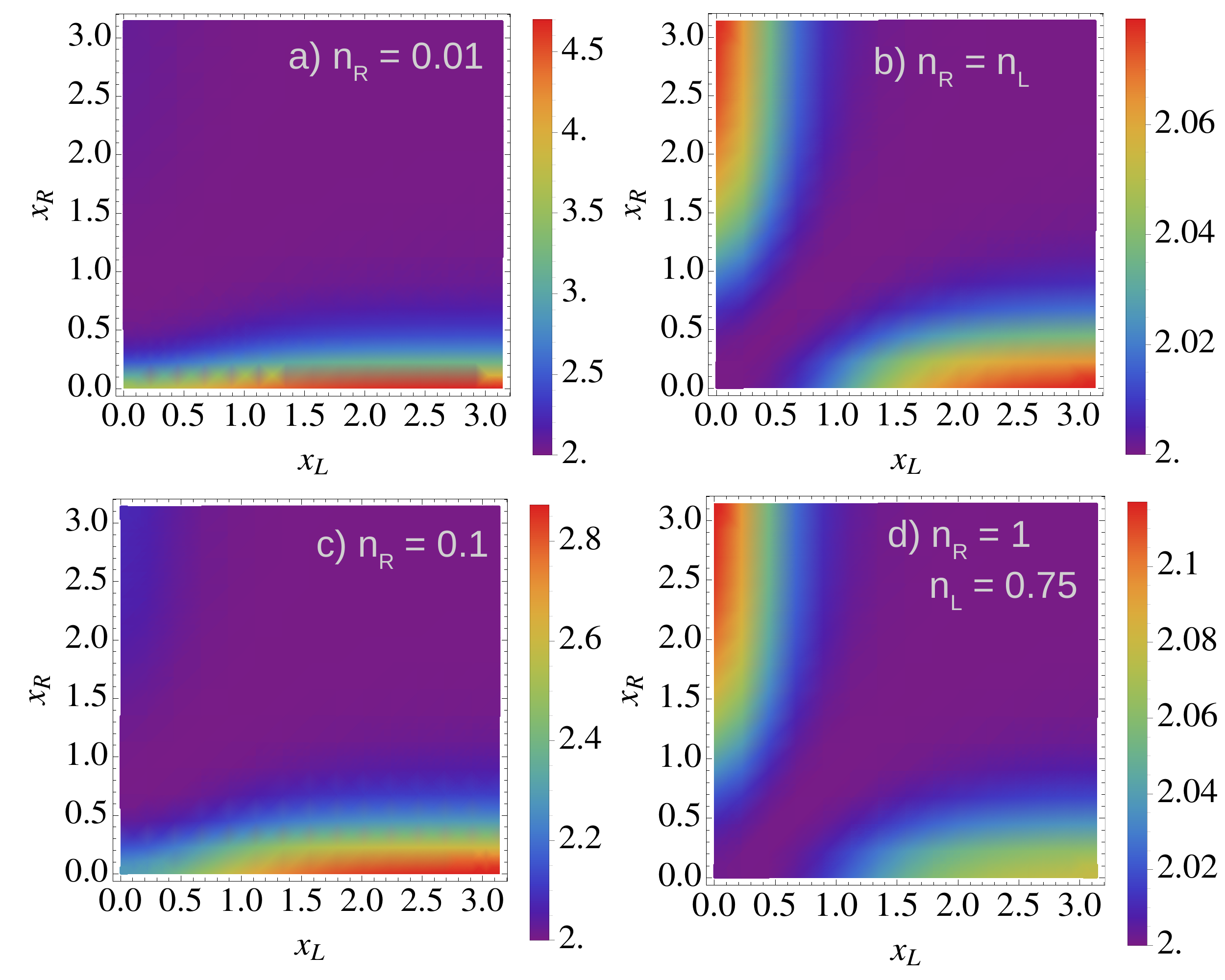}
\caption{Density plots showing the robustness of the thermodynamic uncertainty relationship, $FA\ge 2$ as a function of the squeezing parameters, $x_L$ and $x_R$. Here $n_L=1$ and $n_R=0.1$ in a), b) and c).}
\label{fig-con-FA}
\end{figure}

\begin{figure}
\centering
\includegraphics[width = 0.48 \textwidth]{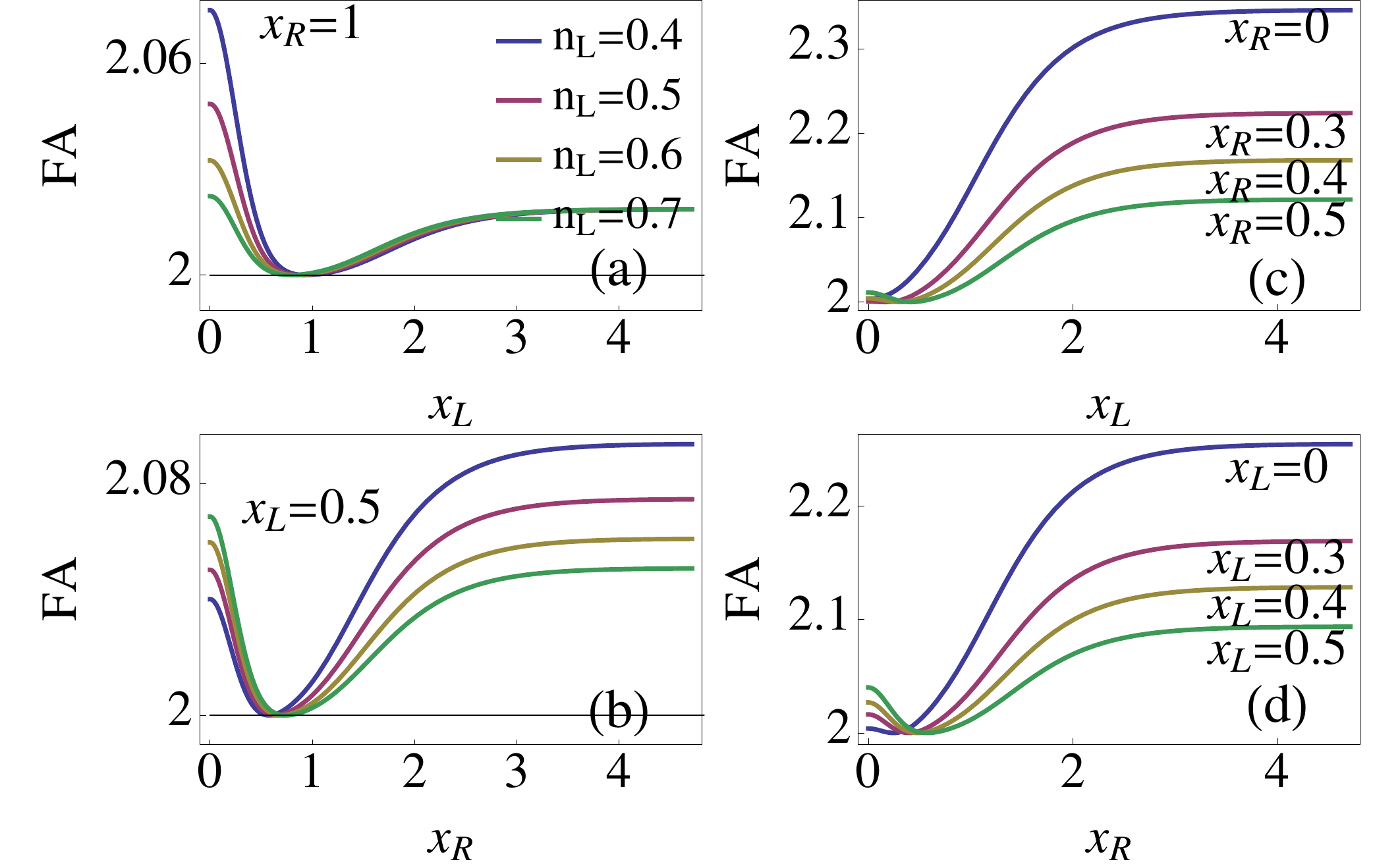}
\caption{Behavior of the product $FA$ as a function of  $x_R$ and $x_R$ for varying $n_L$ in (a), (b). $n_L=0.4,n_R=0.3$ is fixed in (c) and (d). $FA\ge 2$ holds for all the different conditions. }
\label{fig-FA}
\end{figure}

We now focus on the  thermodynamic uncertainty relationship given by $FA>2k_B$, $F$ being the Fano factor, which in our case is given by  
\begin{align}
 F= \frac{j^{(2)}}{j^{(1)}}
\end{align}
When $k_B\to1$, $FA>1$ is shown to be valid for Markovian dynamics in many systems. In our case, we numerically evaluate the quantity $FA$ as a function of $x_L$ and $x_R$ and show the results graphically in Figs.(\ref{fig-con-FA}) and (\ref{fig-FA}). We see that the inequality holds for the entire range of $x_L$ and $x_R$ for a wide range of $n_L$ and $n_R$ values. The robustness is because of the validity of the Gallavoti-Cohen symmetry that leads to a steady state fluctuation theorem. Further it has also been predicted in \cite{saryal2019thermodynamic} that, TURs in bosonic transport holds because of the specific mathematical form of the cumulant generating function \cite{saryal2019thermodynamic}, which makes the internal dynamics of the bath oscillators irrelevant during transport statistics. Such is the case for our system in the Markovian regime, where the analytical expression for the squeezed $S(\lambda)$, Eq.(\ref{S-lam-eq}) and has the same mathematical structure as the one in conventional thermal transport, and hence TUR holds.




\section{Conclusion}
\label{Conclusion}
We theoretically explored the effect of squeezing on the flux and higher order fluctuations during boson exchange between two squeezed harmonic reservoirs and a two level system. Within the Markov approximation, we analytically showed that the direction of the steadystate flux can be reversed under appropriate squeezing conditions. The thermodynamic affinity is shown to be antisymmetric with respect to the interchange of the two squeezing parameters and is non zero when $T_L=T_R$. The odd cumulants are proven to be independent of squeezing when both the left and right reservoirs are squeezed symmetrically. The even cumulants however always depend on the two  squeezing paramaters and are larger in magnitude in comparison to an unsqueezed case. Under high squeezing, the saturation values of the odd and even cumulants can be used to determine the coupling strength and Bose-Einstein distribution of the reservoirs, respectively since these cumulants saturate to two unique values. Under equal temperature setting, the even cumulants behave identically as a function of the squeezing parameters.  A steady state fluctuation theorem is derived  where the thermodynamic affinity is shown to be dependent on squeezing parameters. A  thermodynamic uncertainty relationship holds in the presence of squeezing. 
\begin{acknowledgments}
  MJS acknowledges the project associate fellowship from SERB Grant SERB/SRG/2021/001088. HPG acknowledges the support from the University Grants Commission, New Delhi for the startup research grant, UGC(BSR), Grant No.  F.30-585/2021(BSR).
\end{acknowledgments}  
\section*{appendix}
Upto second order in the perturbation of the coupling, $\hat{V}$, the reduced system dynamics is given by the Eq.(\ref{V-def}):
\begin{align}
\label{eq-V-full}
 \dot{\tilde\rho}=-tr_B\int_o^t ds
(\rho_T(t)\tilde V (t)\tilde V(t_s)-\tilde V(t)\tilde\rho_T(t)\tilde V(t_s)+h.c)
\end{align}
with $t_s:=t-s$, in the interaction picture, $
    \tilde{O}(t)=\exp(i\hat{H}_ot)\hat{O} \exp(-i\hat{H_o}t)$.
$\hat H_o$ contains only the diagonal part from Eq.(\ref{ham-eq}). Assuming that the initial density matrix is factorisable, we can write $\tilde\rho_T:=\tilde\rho\otimes\rho_B$, with
$\rho_B=\rho_L\otimes\rho_R$, being the squeezed density matrices for the left and right reservoirs respectively.
Substituting the value of the operator $\hat V$ in Eq.(\ref{eq-V-full}), we can write the integrands as,
\begin{align}
 \label{eq-vs-1}
 \tilde\rho\rho_B\tilde V_s\tilde V &=
 k_{i\nu}^2\big(\langle\tilde a_{i\nu}^\dag(t_s)\tilde a_{i\nu}(t)\rangle\tilde\rho(\tilde b(t_s)\tilde b(t)+\tilde b(t_s)\tilde b^\dag(t))\\ \nonumber
 &+\langle\tilde a_{i\nu}^{}(t_s)\tilde a_{i\nu}^\dag (t)\rangle\tilde\rho(\tilde b(t_s)\tilde b^\dag(t)+\tilde b^\dag(t_s) b^\dag(t))\big)\\
 \label{eq-vs-2}
 \tilde V\tilde\rho\rho_B\tilde V_s&=
 k_{i\nu}^2\big(\langle\tilde a_{i\nu}^\dag(t)\tilde a_{i\nu}(t_s)\rangle(\tilde b(t)\tilde\rho\tilde b(t_s)+\tilde b(t)\tilde\rho\tilde b^\dag(t))\\ \nonumber
 &+\langle\tilde a_{i\nu}^{}(t)\tilde a_{i\nu}^\dag(t_s) \rangle(\tilde b(t)\tilde\rho\tilde b^\dag(t_s)+\tilde b^\dag(t)\tilde\rho b^\dag(t_s))\big)
\end{align}
where $\langle . \rangle$ represents trace over the squeezed density matrix. On evaluating the matrix elements of Eq.(\ref{eq-vs-1}) and (\ref{eq-vs-2}), $\langle m|.|n\rangle,  m, n =1,0$, only the terms with conjugate system operators survive leading to only $m=n$ terms. The nonzero squeezed bath expectation values are given by \cite{li2017production},
\begin{align}
 \label{eq-bath-cf}
 \langle\tilde a_{i\nu}^\dag\tilde a_{i\nu}\rangle&=\big(\cosh(2x_{i\nu})(n_{\nu}+\frac{1}{2})-\frac{1}{2}\big)f(t,s)\\
 &:=N_\nu f(t,s)\\
 \langle\tilde a_{i\nu}^{}\tilde a_{i\nu}^{\dag}\rangle&=\big(\cosh(2x_{i\nu})(n_{\nu}+\frac{1}{2})+\frac{1}{2}\big)f(t,s)\\
 &:=(1+N_\nu)f(t,s)
 \end{align}
where $n_\nu$ is the Bose function for the $\nu$-th squeezed bath and $f(t,s)= \exp(it_s\omega_{i\nu})\exp(-it\omega_{i\nu})$. With these definitions, a standard Born-Markov approximation ($t\to\infty$) within the wide-band limit we can evaluate the time integrands in Eq.(\ref{eq-vs-1}) and (\ref{eq-vs-2}) by substituting in Eq.(\ref{eq-V-full}). The wide band limit normalizes the squeezing parameter of the $k$th mode of the $\nu$th bath, $x_{i\nu}$ to a real positive number $x_\nu$ \cite{li2017production}. We can now write down two coupled Pauli type master equations with squeezed rates given by,
\begin{align}
    \dot{\rho}_{11}&=-(\Gamma_L(1+N_L)+\Gamma_R(1+N_R))\rho_{11}\\ \nonumber&+ (\Gamma_LN_L+\Gamma_RN_R)\rho_{00}
\\
    \dot{\rho_{22}}&=(\Gamma_L(1+N_L)+\Gamma_R(1+N_R))\rho_{11}\\ \nonumber&- (\Gamma_LN_L+\Gamma_RN_R)\rho_{00}
\end{align}
where $\langle m |\rho 
|m\rangle =\rho_{mm}$ represent the probability of occupation of the occupied and unoccupied states. Redefining the rates as $\alpha_\nu=\Gamma_L(1+N_L)$ and $\beta_\nu=\Gamma_\nu N_\nu$, we can write the above two equations in the Liouville space as,
\begin{align}
 \label{qme}
 |\dot \rho\rangle\rangle &={\breve{\cal L}}
|\rho\rangle\rangle
\end{align}
with the superoperator given by,
\begin{align}
 \breve {\cal L}= \left[\begin{array}{cc}
    -\alpha_L-\alpha_R & \beta_L+\beta_R \\
    \alpha_L +\alpha_R & -\beta_L-\beta_R
\end{array}\right]
\end{align}
Now following the standard procedure of FCS by introducing the auxiliary counting field, $\lambda$ to keep track of the net number of bosons exchanged, $1$ \cite{esposito2009nonequilibrium,harbola2007statistics,goswami2015electron}, we can arrive at Eq.(\ref{lvl-eqn}).

\bibliography{References.bib}

\end{document}